\documentclass[12pt]{iopart}
\usepackage{graphicx}
\usepackage{%epsf,amsmath,amsfonts,
amssymb,amsbsy}
\usepackage[mathscr]{eucal}
\begin{document}

\title{CMBR anisotropy in the framework of cosmological extrapolation of MOND}
\author{V.V. Kiselev}
\address{Russian State Research Center Institute for High Energy Physics,\\
Pobeda 1, Protvino, Moscow Region, 142281, Russia}
\address{Department of Theoretical Physics,\\
Moscow Institute of Physics and Technology (State University),\\
Institutsky 9, Dolgoprudny, Moscow Region, 141701,  Russia}
\eads{\mailto{Valery.Kiselev@ihep.ru}, %%\ead
}

\begin{abstract}
A modified gravity involving a critical acceleration, as empirically
established at galactic scales and successfully tested by data on supernovae
of type Ia, can fit the measured multipole spectrum of anisotropy in the
cosmic microwave background radiation, so that a dark sector of Universe is
constructively mimicked as caused by the dynamics beyond the general
relativity. Physical consequences, verifiable predictions and falsifiable
issues are listed and discussed.
\end{abstract}

 \pacs{98.80.-k, 04.50.Kd, 95.36.+x}

\maketitle

\section{Introduction}

The general relativity as the theory of gravitation is
triumphantly %successfully
tested in ``classical experiments'' \cite{tests} on
\begin{itemize}
  \item the deflection of light by the Sun,
  \item the perihelion precession of Mercury,
  \item the gravitational redshift of electromagnetic radiation,
  \item the time delay of signal from satellites due to the curved
      space-time, and
  \item the gyroscope precession during the orbital motion around the
      Earth as caused by the spatial curvature \cite{testlast,testlast2}.
\end{itemize}
In addition, an expanding Universe is the prediction being inherent for the
general relativity. Sure, the general relativity conceptually is the perfect
theory of classic gravity. In this respect, we usually expect that it is
valid, indeed, until effects of quantum gravity would be essential at
Planckian scales of energy that is unreachable in practice. However, such the
point of view, perhaps, is actually broken:

The general relativity itself gives us a brilliant tool in order to search
for indications, which signalize on breaking down its validity: while
observing a motion by inertia, we get a curvature of space-time, which can be
inserted into Einstein equations, that yields a tensor of energy-momentum for
an appropriate substance, and if properties of the substance are mysterious
and unpredictable, then we get a hint for suspecting of the incorrect
description for the nature. This is exactly the case of hypothetic dark
matter (see, for instance, review in \cite{FMcG}): in the framework of
general relativity, it should be inevitably introduced as a transparent
pressureless substance dynamically isolated from the ordinary visible matter
made of known, well studied particles, except the interaction via the
gravity, so that properties of dark matter are artificially tuned. This
tuning has various aspects.

First, rotational curves in disc-like galaxies, i.e. dependencies of rotation
velocities of stars versus a distance to the galaxy center, if described by
the law of Newtonian gravity, requires the introduction of dark matter with a
tuned spatial distribution. Unexpectedly, the dark matter halo is inessential
in regions, wherein the gravitational acceleration caused by the visible
matter, is greater than a critical value $\tilde g_0$, while the halo starts
to dominate in regions, wherein the acceleration by the visible matter is
less than $\tilde g_0$ \cite{FMcG,rev-MOND1,rev-MOND2,rev-MOND3}. That was M.
Milgrom who first introduced the critical acceleration $\tilde g_0$ in the
description of rotational curves \cite{MOND}. It is spectacular that the
critical acceleration is universal: it does not depend on a genesis of
disk-like galaxy, and it is the same for any studied disc-like galaxy.
Unbelievably, an amount and spatial distribution of dark matter is tuned to
the amount and distribution of visible matter in order to form in cosmic
collisions the dark halos in disc-like galaxies with the same universal
critical acceleration. In the framework of general relativity, there is no
straightforward dynamical mechanism for a deduction of such the universal
acceleration. Anyway, the deduction looks to be very artificial, most
probably, it certainly could be the fine tuning. The critical acceleration
subtly binds the dark matter to the visible matter. If this relation is
dynamical, then it is not due to Newtonian gravity, that does not include any
critical acceleration. Further, in deep regions of dark halo dominance, the
rotation velocities tend to constant values $v_0$, that empirically satisfy
the baryonic Tully--Fisher law \cite{Tully:1977}:
\begin{equation}\label{bTF}
    v_0^4=GM\tilde g_0,
\end{equation}
where $M$ stands for the mass of visible matter in disc-like galaxy, $G$
denotes the Newton constant. Again, the dark matter halo is tuned, so that
the star motion within the halo strictly correlates with the usual visible
matter, while the constant $\tilde g_0$ is universal \cite{McGaugh}. Finally,
in order to complete the first item of argumentation, features in
distributions of visible matter, no doubt,  are imprinted in rotational
curves even in regions of dark matter dominance \cite{FMcG}, hence, features
of dark matter distributions are tuned to the visible matter, though we have
no dynamical reasons for such the correlations in the framework of general
relativity. Moreover, the morphology of spatial distributions is absolutely
different for the baryonic and dark matter in disc-like galaxies: an
exponentially falling central bulge and thin disc of stars and gas in
contrast to power-law decline of dark spherical halo. Thus, the universal
critical acceleration is the mysterious quantity for the general relativity,
that cannot be predicted, while its notion emerged empirically. The critical
acceleration $\tilde g_0$ reveals the fine tuning of hypothetic dark matter
to the ordinary visible matter.

Second, in cosmology with the observed accelerated expansion of Universe by
data on a dependence of brightness of type Ia supernovae versus the red shift
\cite{Riess,SN1,SN2,SN3,SN4}, the general relativity was have to introduce
the extended dark sector, which includes a homogeneous dark energy in
addition to the nonhomogeneous dark matter. In the simplest case, the dark
energy can be represented by the cosmological constant, otherwise it should
be described by a homogeneous fluid $X$ with the state parameter $w_X$ being
the ratio of pressure $p_X$ to energy density $\rho_X$, $w_X=p_X/\rho_X$
close to vacuum value of $-1$, in contrast to the pressureless dark matter
with $w_\mathrm{DM}=0$. Evidently, the nature of dark matter and dark  energy
is very different. But surprisingly, the energy density of dark energy, or
the value of cosmological constant $\Lambda$, is finely tuned to the critical
acceleration \cite{FMcG}, so that
\begin{equation}\label{finet}
    G\rho_X\sim \Lambda \sim \tilde g_0^2.
\end{equation}
Therefore, the dark energy should be inherently connected to the unexpectedly
correlated dynamics of dark matter and ordinary matter. But the coincidence
of (\ref{finet}) is mysterious for the general relativity.

Nevertheless, the general relativity applied to the cosmos still looks
formally viable in the form of concordance model: the ordinary matter
balanced with the cold dark matter (CDM) and cosmological constant $\Lambda$
in the flat space, the $\Lambda\mbox{CDM}$ variant with a spatial curvature
compatible with zero in limits of uncertainties. Moreover, there are two
important successes in the model: 1) a correct fitting for observed
anisotropy of cosmic microwave background radiation (CMBR) \cite{WMAP} that
becomes possible due to a tuned amount of non-baryonic dark matter, and 2) an
appropriate baryon to photon ratio consistent with a current status of big
bang nucleosynthesis (BBN) (see review by Fields B D, Sarkar S in
\cite{PDG}).

Indeed, the anisotropy of CMBR is suppressed by 5 orders of magnitude with
respect to the CMBR temperature, and it is caused by a propagation of sound
waves in a hot photon-electron-baryon medium up to a moment, when the
electrons bind to nuclei to form neutral transparent gas. The snapshot of
Universe at the time of decoupling of photons evolves to us, and it
represents acoustic peaks in the following multipole spectrum of temperature
fluctuations
\begin{equation}\label{spectrum}
    \langle\Delta T(\boldsymbol n_1) \Delta T(\boldsymbol n_2)\rangle=\sum_l
    \frac{2l+1}{4\pi}\,C_l\,P_l(\boldsymbol n_1\cdot\boldsymbol n_2),
\end{equation}
where $\boldsymbol n_{1,2}$ denote directions in the celestial sphere, $P_l$
are Legendre polynomials of multipole number $l$. The spectrum, i.e. $C_l$,
depends on
\begin{itemize}
  \item the Universe evolution,
  \item a primary spectrum of inhomogeneity, and
  \item the propagation of inhomogeneity during the evolution.
\end{itemize}
Then, the Universe evolution is well described by $\Lambda$CMD \cite{WMAP}
and it can be extrapolated to the age of Universe, when the snapshot of CMBR
was done, i.e about 380 thousands years after the big bang. The primary
spectrum of inhomogeneity is suggested to be close to the Harison--Zeldovich
distribution of so called ``no-scale'' limit at a spectral index $n_s(k)=1$:
\begin{equation}\label{HZ}
    \left\langle\frac{\delta\rho^2(0)}{\rho^2}\right\rangle=A \int
    \left(\frac{k}{k_0}\right)^{n_s(k)-1}\,\mathrm{d\ln k},
\end{equation}
where $\delta\rho(\boldsymbol r)=\rho(\boldsymbol r)-\rho$ denotes a contrast
of energy density at comoving coordinate, $k$ is a wave vector conjugated to
the comoving coordinate, $A$ stands for an amplitude at a reference value of
$k_0$. The spectral index and amplitude are subjects to fit the observed
spectrum of $C_l$. Finally, the propagation of inhomogeneity includes the
sound and further smearing of waves by the gravitation. The concordance model
of cosmology in the framework of general relativity well fits $C_l$ with the
flat space and tuned amount of dark components \cite{WMAP}: relative
fractions of dark matter  $\Omega_\mathrm{DM}\approx20$\% and dark energy in
the approximation of cosmological constant $\Omega_\Lambda\approx 76$\%. The
baryonic matter composes only $\Omega_b\approx4$\%. This value is dictated by
heights of distinct initial three acoustic peaks in the multipole spectrum,
$C_l$. That is the dark matter fraction, which regulates the relative heights
and positions of peaks up to small variations due to the parameters of
primary spectrum of inhomogeneity.

Next, the amount of baryonic matter and the temperature of CMBR fixes the
baryon-to-photon ratio of densities $\eta_b=n_b/n_\gamma$, that is the only
free parameter in calculation of elements abundance during the big bang
nucleosynthesis. The current state of measurements of elements abundance
extrapolated to the primary abundance is compatible with $\eta_b$ extracted
from the $\Lambda$CDM fit of CMBR anisotropy \cite{PDG}.

Thus, the success of concordance model stimulates direct searches for an
appropriate heavy dark matter particle at colliders and underground
big-volume experiments, sensitive to suppressed, but non-zero cross sections
of dark matter interaction with the ordinary matter.

However, even a discovery of candidate for the dark matter particle would not
withdraw the problem of ad hoc tuning of dark sector. Moreover, it would
sharpen the need to search for the dynamical reasons causing the adjustment
of dark matter, i.e to look beyond the general relativity.

A model of gravity involving the critical acceleration should naturally
include both empirical laws such as the Tully-Fisher relation at the galactic
scales and correct descriptions of Universe evolution, observed features of
CMBR, large scale structure and elements abundance, so that the model would
give a successful approach being alternative to the general relativity in
cosmology, of course. At galactic scales, M. Milgrom invented the modified
Newtonian dynamics (MOND) \cite{MOND} stating the gravitational acceleration
$\boldsymbol g$
\begin{equation}\label{Milg}
    \boldsymbol g\,\zeta(g/\tilde g_0)=-\boldsymbol \nabla\phi_M,
    \qquad \zeta(y)=\frac{y}{\sqrt{1+y^2}},
\end{equation}
where the critical acceleration extracted from the modern analysis of
rotational curves is given by $\tilde g_0=(1.24\pm0.14)\times
10^{-10}\;m/s^2$ \cite{McGaugh}, while $\phi_M$ denotes the Newtonian
gravitational potential of ordinary matter, satisfying $\boldsymbol \nabla^2
\phi_M=-4\pi\,G\,\rho_M$. At $g/\tilde g_0\gg 1$, we get the Newtonian limit
of gravitational force\footnote{Sub-leading terms are suppressed, so that the
force at the Earth and in the Solar system is not distinguishable from the
Newtonian one.}, while at $g/\tilde g_0\ll 1$ the Tully-Fisher law is
satisfied by construction. Note that (\ref{Milg}) successfully predicts the
rotational curves by the given distribution of visible matter\footnote{The
only parameter of fitting the rotational curves within MOND is the
light-to-mass ratio, which strictly correlates with astrophisical
expectations for given galaxies. Moreover, in gas-rich galaxies this
uncertainty is absent, that means the MOND predicts rotational curves with no
adjustment of any parameters, see details in \cite{FMcG}.} with appropriate
imprints of its features, see review in \cite{FMcG}.

However, the straightforward insertion of (\ref{Milg}) into the dynamics at
cosmological scales would results in an inconsistent distortion of vacuum
homogeneity during the evolution of Universe, for instance, i.e. in the
vacuum instability as was shown in \cite{KT-CQG}, wherein authors offered to
introduce the cosmological behavior of critical acceleration in the form of
\begin{equation}\label{cosmog}
    \tilde g_0\mapsto g_0=g_0^\prime\,|\boldsymbol x|,
\end{equation}
where the distance is determined by comoving coordinate $\boldsymbol r$ and
scale factor of Universe expansion $a(t)$, so that $\boldsymbol
x=a(t)\,\boldsymbol r$. Then, the homogeneous cosmology with the gravity law
modified at accelerations below the critical value of $\tilde g_0$ is
consistent, that constitutes the cosmological extrapolation of MOND
\cite{KT-CQG}. The cosmological regime is matched to MOND at a size of large
scale structure $|\boldsymbol x|_\mathrm{lss}$, i.e. at the characteristic
scale of inhomogeneity, that is related to the acoustic scale and sound
horizon in the baryon-electron-photon plasma (see details in \cite{KT-CQG}).

In the framework of cosmological extrapolation of MOND with the interpolation
function $\zeta(y)$ in (\ref{Milg}) the gravity equations for the evolution
of homogeneous and isotropic Universe can be written in the form
\cite{KT-CQG}
\begin{equation}\label{mg1}
    \left(R_\mu^\nu \xi^\mu \xi_\nu\right)^4=\left(
    \left(R_\mu^\nu \xi^\mu \xi_\nu\right)^2+
    \left(K_\mu^\nu \xi^\mu \xi_\nu\right)^2 \right)
    \left(\bar R_\mu^\nu \xi^\mu \xi_\nu\right)^2,
\end{equation}
in terms of Ricci tensor $R_{\mu\nu}$ for the metric $\eta_{\mu\nu}$ and
4-vector in the direction of cosmological time $\xi^\mu=(1,\boldsymbol 0)$,
wherein the matter energy-momentum tensor $T_{\mu\nu}$ defines
\begin{equation}\label{mg2}
    \bar R_\mu^\nu [\eta]=8\pi G \left(T_\mu^\nu-\textstyle{\frac12} \eta_\mu^\nu
    T\right),
\end{equation}
while the extra tensor of curvature $K_\mu^\nu$ is the Ricci tensor of de
Sitter space-time being both homogeneous in time and space as well as
isotropic
\begin{equation}\label{mg3}
    K_\mu^\nu=3g_0' \eta_\mu^\nu.
\end{equation}
In addition to (\ref{mg1}) the conservation of energy-momentum $\nabla_\mu
T^\mu_\nu=0$ is hold, of course.

 Then, in the limit of general
relativity we put $\left(R_\mu^\nu \xi^\mu \xi_\nu\right)^2\gg
\left(K_\mu^\nu \xi^\mu \xi_\nu\right)^2$, and we find $R_\mu^\nu\approx \bar
R_\mu^\nu$, that results in the Einstein equations
\begin{equation}\label{mg4}
    R_\mu^\nu=8\pi G \left(T_\mu^\nu-\textstyle{\frac12} \eta_\mu^\nu T\right).
\end{equation}
In the limit of $\left(R_\mu^\nu \xi^\mu \xi_\nu\right)^2\ll \left(K_\mu^\nu
\xi^\mu \xi_\nu\right)^2$ we get the modified evolution of Universe,
effective at present.

It is important that parameterizing the size of large scale structure by
$|\boldsymbol x|_\mathrm{lss}\sim \lambda^2/H_0$ at $g_0^\prime\sim
H_0^2/\lambda$ and moderate value of $\lambda\sim \frac17$, we find the
simplest solution for the coincidence problem, because the Milgrom
acceleration $\tilde g_0\sim \lambda H_0$ becomes close to the scale of
cosmological constant $\Lambda\sim H_0^2$. Moreover, we will see that the
dark matter fraction of energy is also regulated by the value of $\lambda$.

\begin{figure}[t]
\begin{center}
\setlength{\unitlength}{0.8mm}
\begin{picture}(130,110)
\put(0,0){  \includegraphics[width=130\unitlength]{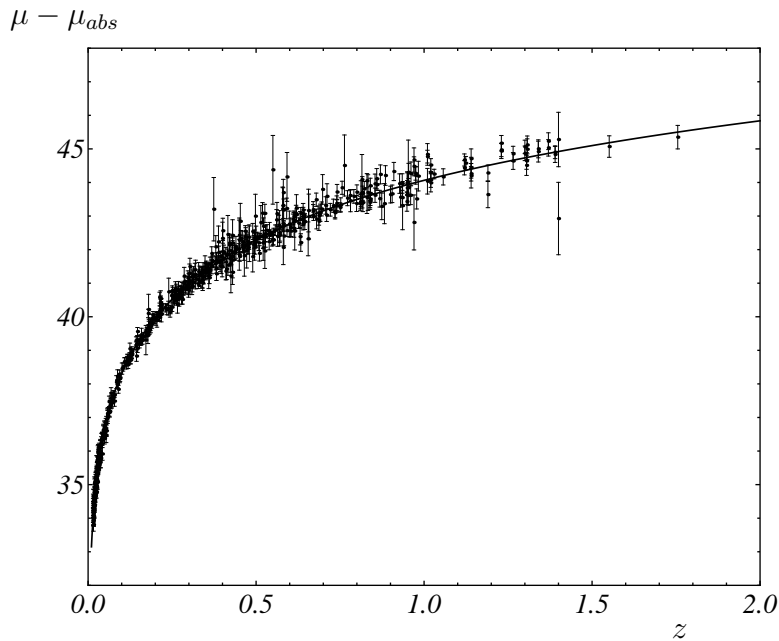}}
\put(110,3){$z$}
\put(0,105){$\mu-\mu_{abs}$}
\end{picture}
\end{center}
  \caption{The magnitude of supernova luminosity versus the redshift $z$. The data with
error bars are taken from the Union2 collection \cite{SN4} and the results of
Hubble Space Telescope on the type Ia supernovae \cite{SN1}. The curve
represents our fit in the framework of modified cosmology.}\label{Uni}
\end{figure}

Eq. (\ref{mg1}) successfully fits the evolution of Universe measured by
observing the brightness of type Ia supernovae versus the red shift
\cite{KT-CQG}. So, the stellar magnitude
\begin{equation}\label{mu}
   \mu=\mu_{abs}+5\log_{10}d_L(z) + 25,
\end{equation}
depends on the photometric distance $d_L$ (in Mpc), determined by the Hubble
constant evolution $H(z)=\dot a/a$,
\begin{equation}\label{dl}
    d_L(z)=(1+z)\int\limits_0^z\frac{dz}{H(z)},
\end{equation}
where $\mu_{abs}$ is an absolute stellar magnitude of light source at the
distance of 10 pc. We show the Hubble diagram for the type Ia Supernovae in
Fig. \ref{Uni}. The mean deviation squared per degree of freedom gives
$\chi^2/\mbox{d.o.f.}=1.03$ for our fit with the following assignment of
parameters:
\begin{equation}\label{q0}
\begin{array}{rcl}
  q_0=-0.853, &\qquad & \,\,z_t=0.375, \\
  \,\,h=0.71, & & \Omega_b=0.115,
\end{array}
\end{equation}
where $q_0$ determines the deceleration parameter at red sift $z=0$,
$q(z)=-\ddot a/aH^2(z)$, $z_t$ stands for the red shift, when the
acceleration is equal to zero, $h$ parameterizes the Hubble rate at $z=0$ via
$H_0=h\cdot 100\,\mathrm{km/s\cdot Mpc^{-1}}$. Accepting the prescription of
\begin{equation}\label{K0}
    g_0^\prime=K_0H_0^2,
\end{equation}
we can find that
\begin{equation}\label{K0-1}
    q_0=\frac12\,K_0\Omega_b\left((1+z_t)^3-1\right),
\end{equation}
when the energy density determined by the cosmological constant, is given by
the energy budget of Universe, $\Omega_\Lambda=1-\Omega_b$. The same values
of deceleration, $q_0$, and red shift of transition from the deceleration to
acceleration, $z_t$, could be obtained in concordance model of
$\Lambda$CDM\footnote{The $\Lambda$CDM parameters are marked by bars.} at
\begin{equation}\label{ratio}
    \frac{\bar\Omega_M}{\bar\Omega_b}=-\frac{K_0}{q_0}.
\end{equation}
Therefore, the modified gravity with the critical acceleration, i.e. at
nonzero $K_0\approx 7.9$, determines the ratio of matter to baryon fractions
of energy, so that at $q_0\sim-1$ we get $\bar\Omega_M/\bar\Omega_b\sim K_0$,
which correlates with the scale of large scale structure, considered above,
$K_0\sim 1/\lambda$.

Note that the quality of fit is not sensitive to valuable variations of
baryonic fraction $\Omega_b$, while it was extracted from the appropriate
value of sound horizon
\begin{equation}\label{sound-h}
    r_s(z)=\int\limits_0^{t(z)}c_s\,dt,
\end{equation}
where the speed of sound is determined by the baryon-to-photon ratio of
energy $R$, so that
\begin{equation}\label{sound}
    c_s=\frac{1}{\sqrt{3}}\,\frac{1}{\sqrt{1+R}},
    \qquad
    R=\frac34\,\frac{\rho_b}{\rho_\gamma}=\frac34\,
    \frac{\Omega_b}{(1+z)\Omega_\gamma}.
\end{equation}
The value of sound horizon at $z=0.2$ and $z=0.35$ was extracted from
baryonic acoustic oscillations (BAO) \cite{BAO}, which, in the case of baryon
matter only, favor for the enhanced estimate of $\Omega_b$ shown above. The
same conclusion follows from the calculation of acoustic scale in the
spectrum of CMBR anisotropy \cite{WMAP},
\begin{equation}\label{la}
    l_A=\frac{\pi d_L(z_*)}{(1+z_*)\,r_s(z_*)},
\end{equation}
here $z_*$ is the redshift of decoupling, when due to the recombination of
electrons with protons the medium becomes transparent for photons (see
analytical approximations for $z_*$ in terms of baryonic density, matter
density and Hubble constant in \cite{HS96}.). WMAP gives $l_A=302.69 \pm
0.76$, while we deduce $l_A=302.5$ \cite{KT-CQG} compatible with the
uncertainty of measurement. Hence, we expect that scale features of CBMR
anisotropy could be fitted in the framework of cosmological extrapolation of
MOND.

In present paper we describe the procedure of fitting the CMBR anisotropy
spectrum with the model of modified gravity and point to accepted
approximations in Section \ref{S2}. Some actual problems associated with the
theory and phenomenology of our model of modified gravity, are considered in
Section \ref{SS2}. We present an itemized discussion of model verification
and falsification in Section \ref{S3}.

\section{Fitting the CMBR anisotropy %by the cosmological extrapolation of MOND
\label{S2}}

The tool for the calculation of multipole spectrum of CMBR anisotropy
\cite{Lewis,Fang,CAMB} operates with the Friedmann equation, which is not
valid in the framework of cosmological extrapolation of MOND. Nevertheless,
we can integrate out the dynamical equations of our model in order to extract
the Hubble rate at any red shift and to parameterize it with a mimic dark
energy contribution in addition to the fraction of baryonic matter. Indeed,
exploring the general relativity in the case of baryonic matter and dark
energy in the isotropic homogeneous curved space we get
\begin{eqnarray}
  \hspace{6mm}\frac{H^2}{H_0^2} &=& \frac{\Omega_b}{a^3}+\frac{\Omega_k}{a^2}+\Omega_X(a), \\[2mm]
  -\frac{2\ddot a/a}{ H_0^2} &=&
  \frac{\Omega_b}{a^3}+\Omega_X(a)(1+3w_X(a)),
\end{eqnarray}
where $\Omega_k$ stands for the contribution of space curvature, and the
energy budget holds $\Omega_b+\Omega_X(1)+\Omega_k=1$. Excluding
$\Omega_X(a)$, we get the expression for the dark equation of state
\begin{equation}\label{darkw}
    w_X(a)=-\frac13\left(1+\frac{2\frac{\ddot a}{aH_0^2}+\frac{\Omega_b}{a^3}}
    {\frac{H^2}{H_0^2}-\frac{\Omega_b}{a^3}-\frac{\Omega_k}{a^2}}\right).
\end{equation}
Here we insert the expression for the acceleration that follows from eq.
(\ref{mg1}) for the modified gravity including the actual values of energy
fractions for baryons and vacuum, $\Omega_b$ and $\Omega_\Lambda$,
respectively,
\begin{equation}\label{ddota}
    \frac{\ddot a}{a H_0^2}=\left(\Omega_\Lambda-\frac{\Omega_b}{2a^3}\right)
    \frac{1}{\sqrt{2}}\sqrt{1+\sqrt{1+\frac{(2g_0^\prime/H_0^2)^2}
    {\left(\Omega_\Lambda-\frac{\Omega_b}{2a^3}\right)^2}}}.
\end{equation}
Note that the red shift of transition from the acceleration to deceleration
of Universe, $z_t$ is related to the fraction of cosmological constant due to
$$
    \Omega_\Lambda=\frac12\,\Omega_b(1+z_t)^3.
$$
Again, eq.(\ref{ddota}) clearly shows that putting $g_0^\prime$ equal to
zero, we get the dynamics of general relativity, otherwise near $\ddot a=0$
the dynamics enters the region of strong dominance of modification and the
greatest deviation from the general relativity, as it is actual at present.

\begin{figure}[bh]
\begin{center}
\setlength{\unitlength}{0.6mm}
\begin{picture}(130,110)
\put(0,0){  \includegraphics[width=130\unitlength]{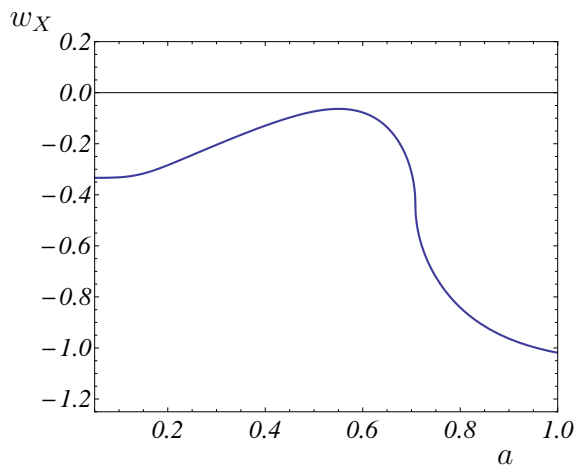}}
\put(110,3){$a$}
\put(2,100){$w_X$}
\end{picture}
\end{center}
  \caption{The equation of state $w_X$ versus the scale factor $a$ in the
  model of modified gravity involving the critical acceleration. }\label{wX}
\end{figure}

Integrating out (\ref{ddota}) at the initial condition $\dot
a/a({t=t_0})=H_0$, we obtain\footnote{In practice, we use the scaling
variable $\tau=tH_0$, that completely covers the differential equations under
consideration.} the scaling quantity $H/H_0$ required for the complete
definition of r.h.s. in (\ref{darkw}). Fig. \ref{wX} shows the behavior of
$w_X$ versus the scale factor $a$ as we have calculated in the cosmological
extrapolation of MOND with parameters listed in (\ref{q0}) at $\Omega_k=0$.
Small variations of the spatial curvature in limits $|\Omega_k|<0.02$,
deceleration parameter $q_0$, transitional red shift $z_t$ and baryonic
fraction $\Omega_b$ within 10\% lead to negligible changes, which are only
just visible in the figure. We emphasize that the modification of gravity
predicts the very specific dependence of equation of state for the dark
energy, that we will discuss in Section \ref{S3}.

After the definition of referenced homogeneous evolution of Universe under
the modified gravity, we can use the standard tool for the calculation of
CMBR anisotropy spectrum \cite{Lewis,Fang,CAMB}. However, in this way the
propagation and smearing of sound waves would be described in the framework
of general relativity with no dark matter, i.e. at $\Omega_\mathrm{DM}=0$,
while we have to modify this procedure in accordance with a structure
formation under the modified law at accelerations below the critical one. The
problem is that such the modification of perturbation transfer function is
not linear, and the appropriate machinery of calculation is not yet
developed. That is missing point of our consideration, of course.
Nevertheless, the extensive usages of tool have shown that the influence of
dark matter on the spectrum is reduced to relative enhancement of third peak,
whereas this enhancement is due to enforcing the gravity. So, since the
modified gravity produces the very similar effect of enforcing the gravity,
we can expect that it could results in the same fine feature as concerns for
the enlarging the third peak.

In this respect, we have to emphasize that a complete axiomatic approach
based on a formulation of action for a modified gravity in terms of given
extended set of gravitational fields has got the advantage in calculating of
various predictions including the CMBR anisotropy. So, we can mention the
following fully relativistic schemes (more examples and references find in
\cite{FMcG}):
\begin{itemize}
\item Bekenstein's theory of tensor-vector-scalar (TeVeS) gravitational
    fields \cite{TVS} involving Maxwellian vector field and reproducing
    the critical acceleration in the case of isolated gravitational
    source, equivalent to MOND,
\item Moffat's modified gravity \cite{Moffat,Moffat2,MT}, giving the
    approximation of gravity law similar to the MOND,
\item generalized TeVeS theories \cite{Skordis:2008pq} with
    non-Maxwellian vector field.
\end{itemize}
However, first, these theories include dark gravitational fields actually
replacing the dark matter that looks like a refinement of problem. Second,
all of them have the strict theoretical illness: there are configurations
with unlimited, infinite negative energy, that leads to instability of
physically sensible solutions (see details and references in \cite{FMcG}).
Third, the critical acceleration is still introduced ad hoc with no
reasonable connection to the present Hubble rate or the cosmological
constant.

So, we prefer for the phenomenological approach, which does not introduce new
artificial and heuristic notions. In this way, we can investigate the role of
critical acceleration in the modified gravity by studying various phenomena
step by step in order to find fundamental features and differences from the
general relativity.

At present, we try to fit the spectrum of CMBR anisotropy by using the
modified evolution of homogeneous Universe and optimizing the primary
spectrum of inhomogeneity, which develops as the sound smeared by ordinary
gravity. So, we suggest that a modification of smearing will not be very
crucial for the main features of spectrum.

\begin{figure}[b]
\begin{center}
\setlength{\unitlength}{0.9mm}
\begin{picture}(130,110)
\put(0,0){  \includegraphics[width=130\unitlength]{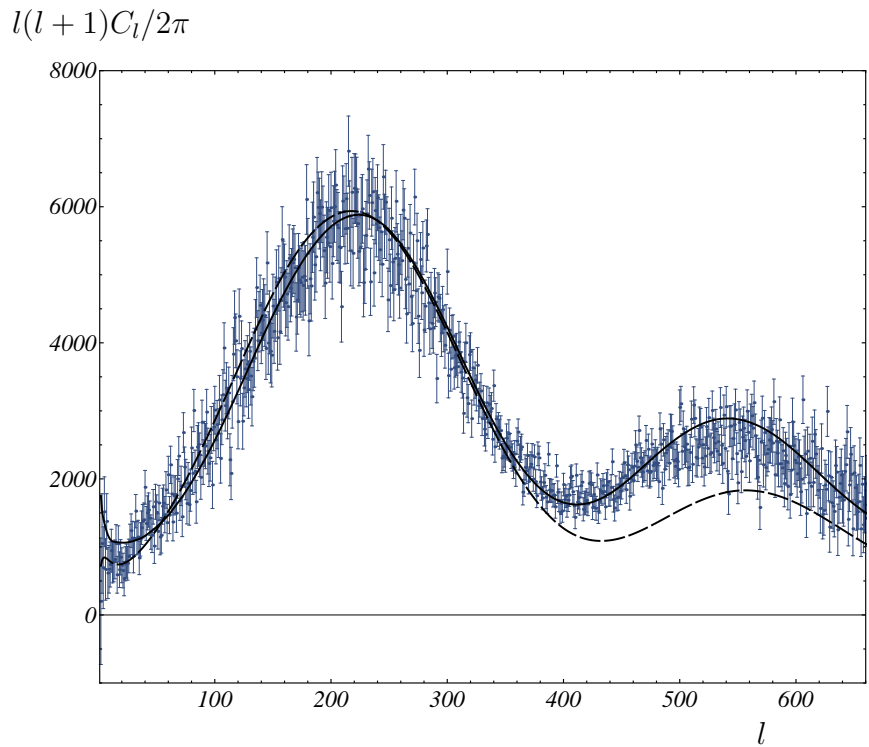}}
\put(110,0){$l$}
\put(0,105){$l(l+1)C_l/2\pi$}
\end{picture}
\end{center}
  \caption{The spectrum  of CMBR anisotropy calculated in the model of modified gravity,
  $l(l+1)C_l/2\pi$ (in $\mu K^2$)
  versus the multipole number $l$, in comaprison with WMAP data \cite{WMAP}.
  The dashed line represents the Harrison--Zeldovich approximation, while
  the solid curve corresponds to the running spectral index (as described in the text).
    }\label{spectr1}
\end{figure}

\begin{figure}[p]
\begin{center}
\setlength{\unitlength}{1.05mm}
\begin{picture}(130,95)
\put(0,3){  \includegraphics[width=130\unitlength]{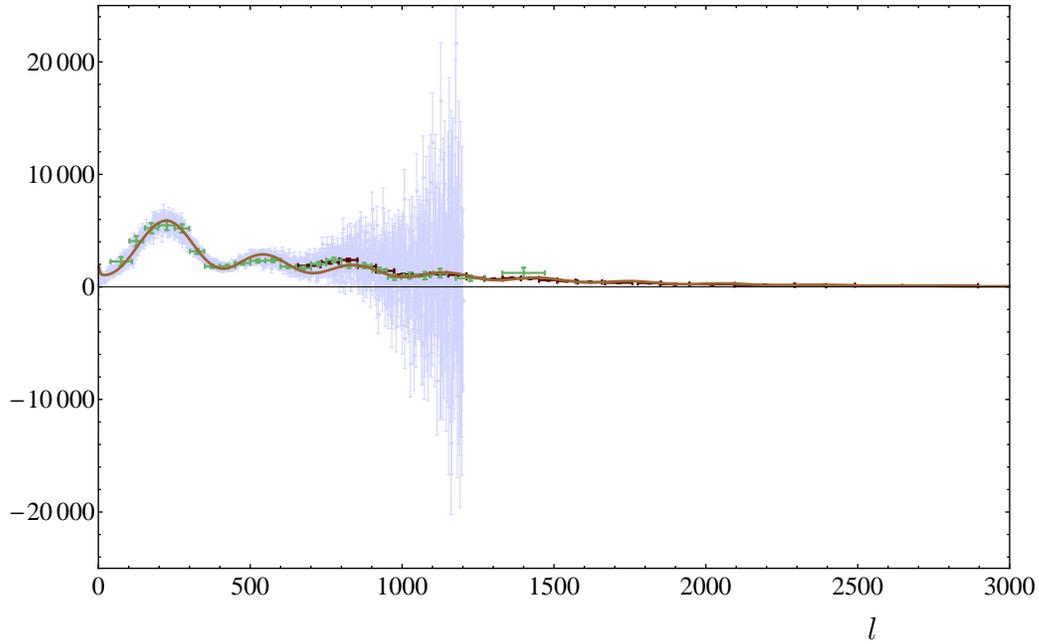}}
\put(110,-2){$l$}
\put(0,87){$l(l+1)C_l/2\pi$}
\end{picture}
\end{center}
  \caption{The same as in Fig. \ref{spectr1} with addition of BOOMERANG data
  \cite{Boom,Boom2} (light crosses) and ACBAR data \cite{Acbar} (dark crosses).
    }\label{spectr2}
\end{figure}

\begin{figure}[p]
\begin{center}
\setlength{\unitlength}{1.05mm}
\begin{picture}(130,87)
\put(0,1){  \includegraphics[width=130\unitlength]{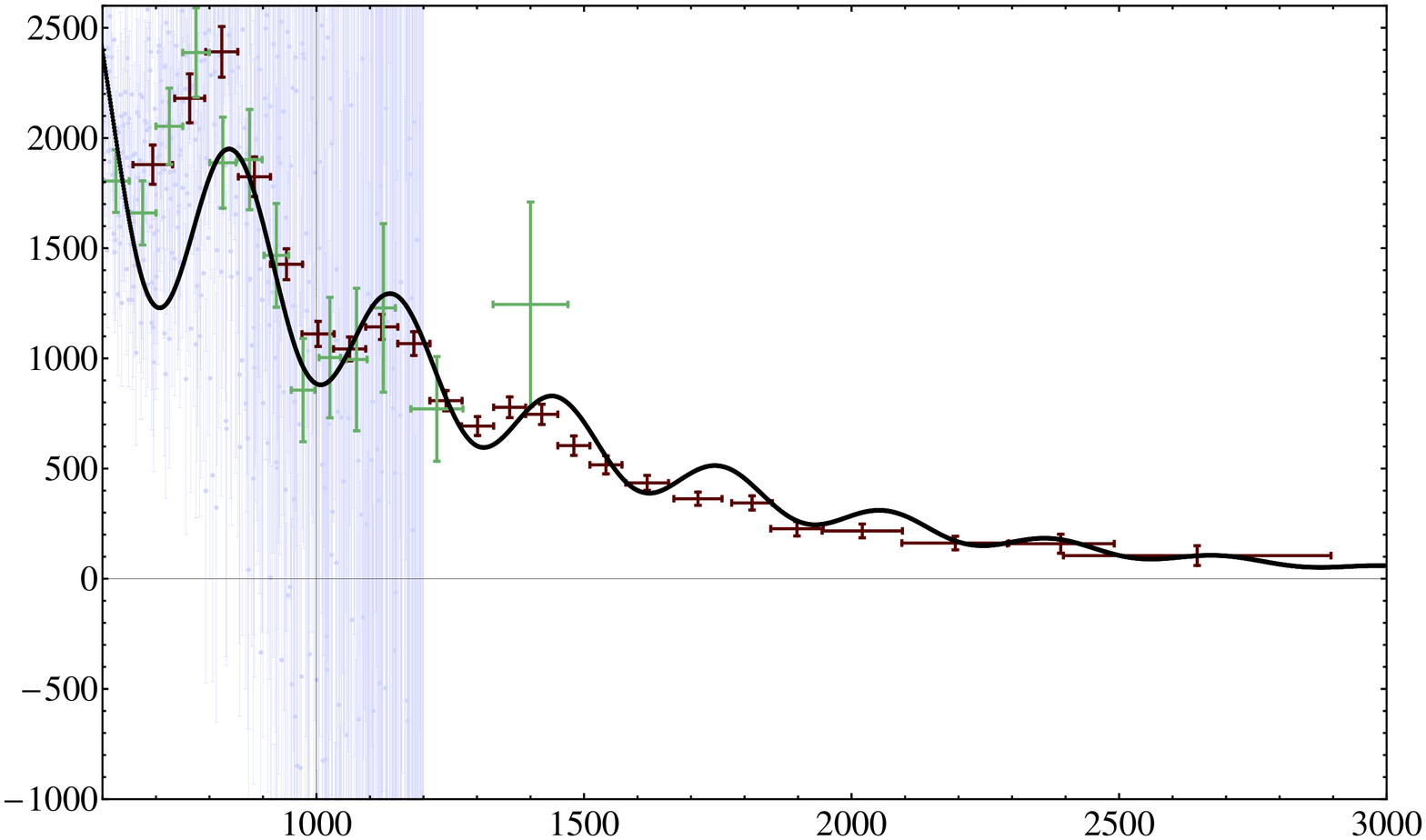}}
\put(110,-2){$l$}
\put(0,82){$l(l+1)C_l/2\pi$}
\end{picture}
\end{center}
  \caption{The same as in Fig. \ref{spectr2}, but focused at the region of
  third and further peaks ($l>600$).
    }\label{spectr3}
\end{figure}

The results of such the fitting are presented in Figs.
\ref{spectr1}--\ref{spectr3}. First, we study the fit of WMAP data
\cite{WMAP} with the Harrison--Zeldovich primary spectrum (HS) at $n_s=1$,
which is shown in Fig. \ref{spectr1} by dashed line. It is spectacular that
the no-scale HS prescription correctly reproduces the angular scale of
multipole momentum, i.e. the position and profile of first acoustic peak.
This feature is obtained due to the correct adjustment of this scale by the
sound horizon in the case of no dark matter \cite{KT-CQG} as mentioned in the
Introduction. Then, we find that the running of spectral index
$n_s(k)=n_s^{(0)}+n_s^\prime\ln k/k_0$ leads to suitable description of both,
first and second acoustic peaks in the modified cosmology with no dark
matter. Setting parameters equal to the following values\footnote{We list the
quantities, which we change from default values assigned in the tool
\cite{Lewis,Fang,CAMB}. Other quantities have been set to its standard
prescriptions.}:
\begin{equation}\label{cosmoMOND}
    \begin{array}{rclcrcl}
       n_s^{(0)} & = & 1.625, & & n_s^\prime & = & 0.24,  \\
       A & = & 5.5\times 10^{-9}, & & \Omega_k & = & -0.015,  \\
       z_{re} & = & 23, & & k_0 & = & 0.04\; \mbox{Mpc}^{-1},  \\
     \end{array}
\end{equation}
we get the fit shown by solid line in Fig. \ref{spectr1}. Set
(\ref{cosmoMOND}) needs comments.

First, the running of spectral index allows to adjust the relative height of
second acoustic peak in the spectrum\footnote{The opportunity to fit the
second peak in the model with suppressed dark matter was considered in
\cite{second-McG}.}. This running is dynamically essential and numerically
significant. Moreover, because of sizable value of slope of spectral index,
we expect that the approximation neglecting the higher orders of expansion
versus the logarithm of comoving wave vector $\ln k$, would be inaccurate at
large intervals of multipole moment.

Second, we introduce a small spatial curvature in order to adjust the
position of peaks, which have been displaced under the strong running of
spectral index. Then, the spatial curvature essentially improves the quality
of fit. Remember, that such the low value of spatial curvature is consistent
with the equation of state we have deduce from the modified
gravity\footnote{Note that at $a\to 0$ the equation of state for the mimic
dark energy tends to $w_X(0)\approx -\frac13$, which points to a possibility
of nonzero spatial curvature.}. Moreover, acceleration $\ddot a$ does not
involve the spatial curvature because of its specific value of state
parameter $w_k=-\frac13$, giving $\rho_k+3p_k=0$. We note also that a nonzero
value of spatial curavture obeys the scaling $|K_0^2\Omega_k|\sim 1$, that is
a feature consistent with our solution or treatment of coincidence problem
(see the Introduction).

Third, the red shift of reionization $z_{re}$, when inhomogeneities of cold
gas are contacted due to the gravity in order to form stars, refers to the
heating of gas, that transforms it to plasma again. It causes a reduction of
intensity of radiation passing through the hot secondary plasma. So, the
amplitude of primary CMBR, $A$ correlates with the red shift of reionization.
In MOND with the enhanced gravity at the stage of diluted gas, one expects
that the star formation starts early than in the general relativity
\cite{FMcG}, hence, we try to improve the fit quality by enlarging the red
shift of reionization\footnote{We set $z_{re}$ via ``a sizable change'': the
typical value of $z_{re}=11$ in $\Lambda$CDM has been enlarged twice for a
distinguishability with no strict reasons or prerequisites.}. The
corresponding improvement of mean deviation squared is equal to -0.11 per
degree of freedom. The optical depth is shifted from $0.9$ to $0.78$. So, we
refer this adjustment as fine effect beyond a sole significance, as well as
slow variation of pivot wave vector $k_0$ from $0.05$ to
$0.04\;\mbox{Mpc}^{-1}$.

Thus, in the simplest way of modifying the background cosmology by the
cosmological extrapolation of MOND, we find that WMAP data \cite{WMAP} can be
fitted at the following mean deviation squared per degree of freedom:
\begin{equation}\label{chi2}
    \chi^2/\mbox{d.o.f.}=1.34,
\end{equation}
which is just 0.19 worse than the typical value of $\chi^2/\mbox{d.o.f.}$ in
the concordance model of $\Lambda$CDM. Fig. \ref{spectr2} clearly shows that
at $l>600$ the WMAP data suffer from huge statistical and systematical
uncertainties. Indeed, the inclusion or exclusion of data at $l>600$, in
practice, do not change $\chi^2/\mbox{d.o.f.}$ Therefore, the WMAP data
beyond the first and second peaks are not conclusive, at all to the moment.

Then, higher multipoles can be studied due to the BOOMERANG and ACBAR data
sets \cite{Boom,Boom2,Acbar}. We present the comparison in Fig.
\ref{spectr3}. Evidently, the dark matter indication is reduced to the
different form of third peak rise. Therefore, we can hope that a nonlinear
smearing of sound waves in MOND could give the same effect in the region of
third peak, too, although to the moment we see the clear tension between the
data in the region of third acoustic peak and the simplest version of fit
based on the modified cosmology of homogeneous Universe only. Thus, we need
the development of procedure to calculate the transfer function of
inhomogeneity versus the red shift \cite{EH97} in the case of nonlinear MOND.
The same is true as concerns for the simulation of large scale structure
formation visible at present time.

Nevertheless, we insist on quite the successful fitting of CMBR anisotropy in
the framework of cosmological extrapolation of MOND, as the variant of
modified gravity involving the critical acceleration in the case of
homogeneous matter.

\section{Main problems\label{SS2}}

The enhanced value of baryon fraction in the energy budget, $\Omega_b\approx
0.115$ implies the enhanced value of baryon-to-photon ratio $\eta_b$ being
the only free parameter of BBN calculation \cite{PDG}. The BBN takes place
during the period when the deviations from the general relativity are
negligible. Therefore, the primary abundance of light elements in
cosmological extrapolation of MOND should differ from the BBN estimated
within the concordance model of $\Lambda$CDM.

At present the data on the helium abundance has large uncertainties which are
compatible with both models under consideration. Then, the deuterium and
lithium primary abundances are clearly able to discriminate between the
enhanced value of $\eta_b$ and its concordance value. However, these
quantities are not measured directly, they are extrapolations from a
suggested model of evolution.

Indeed, we observe the visible sources of light, that mean the primary matter
is contracted in stars with further development of nuclear reactions, and a
model of star evolution has to reproduce the primary values of abundances. In
this respect, one has to take into account the different red shifts of star
formation, i.e. different ages of luminous objects, as considered in the
framework of general relativity or MOND, of course. Next, at present the
standard model for the extrapolation to primary abundances is not self
consistent, because it predicts different ratios of both deuterium to
lithium-7 and lithium-7 to lithium-6, whereas the former ratio is in bright
tension of prediction with the extracted values, while the later is in a deep
contradiction (about three order of magnitude!). Thus, the present BBN status
can not be surely conclusive. It is desirable to get more better reliability
of procedure for the empirical extraction of primary abundance of light
elements.

On the other hand, the doubling of baryonic fraction should appear in
cosmological effects. So, this doubling could appropriately explain a missing
mass in galactic clusters, as found even within MOND. Then, a significant
portion of baryonic matter should be in cold form (Jupiter-like objects), for
instance. Note that even in $\Lambda$CDM the visible matter composes the
tenth fraction of all baryonic matter, only, hence, the most of baryons are
in cold form.

Next, the visible large scale structure should be explained by appropriate
propagation of primary spatial inhomogeneities. However, this issue of
modified gravity involving the critical acceleration is not yet developed
because of nonlinearity of the problem.

\section{Discussion and conclusion\label{S3}}

In this paper, we have shown that the cosmological extrapolation of MOND as
the modified gravity involving the critical acceleration, can successfully
reproduce main features of multipole spectrum of CMBR anisotropy.

Let us list conclusions of our investigation.
\begin{enumerate}
  \item The modified dynamics adjusted to empirical values of the Hubble
      rate, sound horizon in the baryon-photon medium, acoustic scale in
      the multipole spectrum of CMBR anisotropy and magnitudes of type Ia
      supernovae at red shifts $z<2$, mimics the dark energy with the
      very specific \textit{equation of state} $w_X$, shown in Fig.
      \ref{wX}. This is the \textit{falsifiable} prediction of
      cosmological extrapolation of MOND. It can be verified in the
      nearest future by extensive measuring of type Ia supernova
      magnitudes versus the red shift. Even $z<2$, i.e. the scale factor
      variation within the interval $0.3< a <1$, would be enough in order
      to make decision on the direct falsification of cosmological model
      involving the critical acceleration. Moreover, such the exotic
      behavior of dark energy state parameter $w_X$, if would be
      confirmed, will be marginally artificial for the general
      relativity, that would mean the straightforward indication of its
      inadequateness.
  \item The \textit{spectral index} of primary spatial inhomogeneity
      $n_s(k)=n_s^{(0)}+n_s^\prime\ln k/k_0$, essentially runs.
\begin{itemize}
  \item The character of running is \textit{model-dependent}, and
  \item It is very different in the concordance model of general
      relativity and in the cosmological extrapolation of MOND: in
      the modified gravity the running is rather fast, and its
      parameters signalize on the hybrid (multified) inflation, that
      could generate such the spectrum, while the general relativity
      gives the slow running, which preferably corresponds to the
      simplest top-hill inflation due to a single inflaton field
      \cite{inflation}.
  \item The modified gravity gives $n_s^{(0)}>1$ at
      $n_s^\prime\approx\frac14>0$, while the general relativity
      results in $n_s^{(0)}<1$ at $n_s^\prime\to 0$.
  \item The fast running probably indicates the need to improve the
      calculation tool in order to include higher derivatives of
      spectral index with respect to logarithm of wave vector.
\end{itemize}
  \item The spectral index of primary inhomogeneity define initial
      conditions for the transfer of inhomogeneity during the evolution,
      which can be observed in baryonic acoustic oscillations in large
      scale structure of present Universe \cite{BAO}. The procedure of
      calculating the transfer function in the framework of modified
      gravity with the critical acceleration is nonlinear, and it is not
      still developed, that does not allow us to make a comparison with
      data, at present. The structure growth is enhanced in MOND, and it
      can probably need for additional mechanism of smearing the acoustic
      oscillations \cite{FMcG}.
  \item The modified gravity in our version results in doubly enhanced
      fraction of baryons, approximately. This means that
      baryon-to-photon ratio is twice large, at least.
\begin{itemize}
  \item More reliable estimates of primary \textit{abundance of light
      elements} is required in order to \textit{discriminate} the
      general relativity from the modified gravity by the big bang
      nucleosynthesis. Therefore, \textit{BBN} can give the
      \textit{falsification} of cosmological extrapolation of MOND.
  \item The doubling of baryons in the form of cold baryonic matter
      (Jupiter-like objects, for instance) should be found in
      observations. For instance, the mass deficit in galaxy clusters
      described within MOND, can signalize on the appropriate
      enhancement of baryons.
\end{itemize}
  \item The multipole spectrum of CMBR anisotropy needs improvements of
      accuracy in the range of third acoustic peak. If the model of
      modified gravity will still miss the correct description of third
      peak after such the improvement, then this would point to the
      extension of simplest consideration by strict inclusion of
      inhomogeneity propagation within the modified dynamics.
\end{enumerate}

Finally, we have shown that the coincidence problem of general relativity is
inherently solved in the framework of cosmological extrapolation of MOND: the
critical acceleration is connected to the extra Ricci tensor of de Sitter
space, involved in the gravity equations; then, it is naturally correlates
with the cosmological constant. In addition, the modified gravity is mostly
effective at zero acceleration of Universe expansion. That is why the
coincidence notion is actual at present.

Keeping in mind soluble problems mentioned, we state that the cosmological
extrapolation of MOND is quite successful in cosmology. Moreover, we can
falsify it in the nearest future, although the same note on the verification
is actual for the general relativity, too.

\ack%%%nowledgments

This work was partially supported by the grant of Russian Foundations for
Basic Research 10-02-00061. I thank Dr. Timofeev S A for discussions.


\section*{References}
\begin{thebibliography}{99}
\bibitem{tests} Will C M 2005
% The confrontation between general relativity and experiment.
\emph{Living Rev. Rel.} {\bf 9} 3 %-100 (2005).
(\emph{Preprint} gr-qc/0510072])
%%CITATION = GR-QC/0510072;%%
\bibitem{testlast} Everitt F et al. 2008 in The Gravity Probe B experiment.
    Science results - NASA final reports
    {\footnotesize
    (http://einstein.stanford.edu/content/final\_report/GPB\_Final\_NASA\_Report-020509-web.pdf)}
\bibitem{testlast2}
	Everitt C W F  et al. 2011
 %``Gravity Probe B: Final Results of a Space Experiment to Test General Relativity,''
\emph{  Phys.\ Rev.\ Lett.\ } {\bf 106} 221101 %(2011)
(\emph{Preprent} 1105.3456 [gr-qc])
  %%CITATION = ARXIV:1105.3456;%%
\bibitem{FMcG}  Famaey B, McGaugh S 2011
  {Modified Newtonian Dynamics: A Review}. \emph{Preprint}
  arXiv:1112.3960 [astro-ph.CO].
  %%CITATION = ARXIV:1112.3960;%%
\bibitem{rev-MOND1}
 Milgrom M 2011
  {MD or DM? Modified dynamics at low accelerations vs dark matter}
  \emph{Preprint} 1101.5122 [astro-ph.CO]
  %%CITATION = ARXIV:1101.5122;%%
\bibitem{rev-MOND2} Milgrom M 2008 {The MOND paradigm} \emph{Preprint}
    0801.3133
    [astro-ph]
%%CITATION = ARXIV:0801.3133;%%
\bibitem{rev-MOND3} Milgrom M 2009 {MOND: time for a change of mind?}
    \emph{Preprint} 0908.3842 [astro-ph.CO]
%%CITATION = ARXIV:0908.3842;%%
\bibitem{MOND} Milgrom M 1983
%\emph{A Modification Of The Newtonian Dynamics
%    As
%    A Possible Alternative To The Hidden Mass Hypothesis,}
\emph{Astrophys.\  J.\ }    {\bf 270} %(1983)
    365
%%CITATION = ASJOA,270,365;%%
\bibitem{Tully:1977}
  Tully R B, Fisher J R 1977
  %``A New method of determining distances to galaxies,''
  \emph{Astron.\ Astrophys.\ } {\bf 54} 661 %(1977).
  %%CITATION = AAEJA,54,661;%%
\bibitem{McGaugh} McGaugh S S 2011
%\emph{A Novel Test of the Modified Newtonian
%    Dynamics with Gas Rich Galaxies,}
\emph{Phys.\ Rev.\ Lett.\  }{\bf 106}  121303
  (\emph{Preprint} 1102.3913 [astro-ph.CO])
%%CITATION = ARXIV:1102.3913;%%
\bibitem{Riess}
  Riess A G et al.  [Supernova Search Team Collaboration] 2004
%  \emph{Type Ia Supernova Discoveries at z>1 From the Hubble Space Telescope:
%  Evidence for Past Deceleration and Constraints on Dark Energy Evolution,}
  \emph{Astrophys.\ J.\  }{\bf 607} 665
  (\emph{Preprint} astro-ph/0402512)
  %%CITATION = ASJOA,607,665;%%
\bibitem{SN1}
 Riess A G  et al. 2007
%  \emph{New Hubble Space Telescope Discoveries of Type Ia Supernovae at $z > 1$:
%  Narrowing Constraints on the Early Behavior of Dark Energy,}
    \emph{
  Astrophys.\ J.\  }{\bf 659} 98
  (\emph{Preprint} astro-ph/0611572)
  %%CITATION = ASJOA,659,98;%%
\bibitem{SN2}
  Astier P et al.  [The SNLS Collaboration] 2006
%  \emph{The Supernova Legacy Survey: Measurement of $\Omega_M$, $\Omega_\Lambda$ and
%  $w$
%  from the First Year Data Set,}
\emph{
  Astron.\ Astrophys.\  }{\bf 447} 31
  (\emph{Preprint} astro-ph/0510447)
  %%CITATION = AAEJA,447,31;%%
\bibitem{SN3}
  Wood-Vasey W M et al.  [ESSENCE Collaboration] 2007
%  \emph{Observational Constraints on the Nature of the Dark Energy: First
% Cosmological Results from the ESSENCE Supernova Survey,}
 \emph{
  Astrophys.\ J.\  }{\bf 666} 694
  (\emph{Preprint} astro-ph/0701041)
  %%CITATION = ASJOA,666,694;%%
\bibitem{SN4} Amanullah  R et al. 2010
%  \emph{Spectra and Light Curves of Six Type Ia Supernovae at $0.511 < z < 1.12$ and
%  the Union2 Compilation,}
\emph{
  Astrophys.\ J.\  }{\bf 716} 712
  (\emph{Preprint} 1004.1711 [astro-ph.CO])
  %%CITATION = ASJOA,716,712;%%
\bibitem{WMAP} Komatsu E et al.  [WMAP Collaboration] 2010
%    \emph{Seven-Year
%    Wilkinson Microwave Anisotropy Probe (WMAP) Observations: Cosmological
%    Interpretation,}
\emph{Astrophys.\ J.\ Suppl.\  }{\bf 192}  18
    (\emph{Preprint} 1001.4538 [astro-ph.CO])
%%CITATION = APJSA,192,18;%%
\bibitem{PDG}
  Nakamura K { et al.}  [Particle Data Group Collaboration] 2010
  %``Review of particle physics,''
\emph{  J.\ Phys.\ G}  {\bf 37} 075021 %(2010).
  %%CITATION = JPHGB,G37,075021;%%
\bibitem{KT-CQG}
  Kiselev V V, Timofeev S A 2012
%  Cosmological extrapolation of modified Newtonian dynamics.
  \emph{Class.\ Quant.\ Grav.\ } {\bf 29} 065015
  (\emph{Preprint} 1104.3654 [gr-qc])
  %%CITATION = ARXIV:1104.3654;%%
\bibitem{BAO}
  Percival W J, Cole S, Eisenstein D J, Nichol R C, Peacock J A, Pope A C and
  Szalay A S 2007
%  \emph{Measuring the Baryon Acoustic Oscillation scale using the SDSS and
%  2dFGRS,}
\emph{
  Mon.\ Not.\ Roy.\ Astron.\ Soc.\  }{\bf 381} 1053
  (\emph{Preprint} 0705.3323 [astro-ph])
  %%CITATION = MNRAA,381,1053;%%
\bibitem{HS96}
    Hu W and Sugiyama N 1996
%  \emph{Small scale cosmological perturbations: An Analytic approach,}
\emph{  Astrophys.\ J.\  }{\bf 471} 542
  (\emph{Preprint} astro-ph/9510117)
  %%CITATION = ASJOA,471,542;%%
\bibitem{Lewis}
  Lewis A, Challinor A and Lasenby A 2000
  %``Efficient computation of CMB anisotropies in closed FRW models,''
\emph{  Astrophys.\ J.\ } {\bf 538} 473 %(2000)
(\emph{Preprint} astro-ph/9911177)
  %%CITATION = ASTRO-PH/9911177;%%
\bibitem{Fang}
  Fang W, Hu W and Lewis A 2008
  %``Crossing the Phantom Divide with Parameterized Post-Friedmann Dark Energy,''
\emph{  Phys.\ Rev.\ D} {\bf 78} 087303 %(2008)
(\emph{Preprint}  0808.3125 [astro-ph])
  %%CITATION = ARXIV:0808.3125;%%
\bibitem{CAMB} Code for anisotropies in the microwave background (CAMB)
    {\footnotesize http://camb.info/}
\bibitem{TVS} Bekenstein J D 2004
%  \emph{Relativistic gravitation theory for the MOND paradigm,}
\emph{  Phys.\ Rev.\  D} {\bf 70} 083509
  [Erratum 2005 \emph{ibid}   {\bf 71} 069901 ]
  (\emph{Preprint} astro-ph/0403694)
  %%CITATION = PHRVA,D70,083509;%%

\bibitem{Moffat} Moffat J W 2006
%  \emph{Scalar-tensor-vector gravity theory,}
  \emph{JCAP }{\bf 0603} 004
  (\emph{Preprint} gr-qc/0506021)
  %%CITATION = JCAPA,0603,004;%%

\bibitem{Moffat2}
 Moffat J W  2011
  {Modified Gravity or Dark Matter?}
  \emph{Preprint} 1101.1935 [astro-ph.CO]
  %%CITATION = ARXIV:1101.1935;%%

\bibitem{MT}
  Moffat J W and Toth V T 2011
  {Cosmological observations in a modified theory of gravity (MOG)}
  \emph{Preprint} 1104.2957 [astro-ph.CO]
  %%CITATION = ARXIV:1104.2957;%%

\bibitem{Skordis:2008pq}
  Skordis C 2008
  %``Generalizing tensor-vector-scalar cosmology,''
  \emph{Phys.\ Rev.\ D} {\bf 77} 123502
  (\emph{Preprint} 0801.1985 [astro-ph])
  %%CITATION = ARXIV:0801.1985;%%
\bibitem{second-McG} McGaugh S 2000 \emph{Astrophys.\ J.} {\bf 541} L33

\bibitem{Boom} Ruhl J E, Ade P A R, Bock J J, Bond J R,
    Borrill J, Boscaleri A, Contaldi C R and Crill  B P { et al.} (2003)
  %``Improved measurement of the angular power spectrum of temperature
  %anisotropy in the CMB from two new analyses of BOOMERANG observations,''
\emph{  Astrophys.\ J.\ } {\bf 599} 786 %(2003)
(\emph{Preprint}  astro-ph/0212229)
  %%CITATION = ASTRO-PH/0212229;%%
\bibitem{Boom2}
 Jones W C, Ade P, Bock J, Bond J, Borrill J, Boscaleri A, Cabella P and Contaldi C { et
 al.} 2006
  %``A Measurement of the angular power spectrum of the CMB
  %temperature anisotropy from the 2003 flight of BOOMERANG,''
\emph{  Astrophys.\ J.\ } {\bf 647} 823 %(2006)
(\emph{Preprint}   astro-ph/0507494)
  %%CITATION = ASTRO-PH/0507494;%%

\bibitem{Acbar}
  Kuo C { et al.}  [ACBAR Collaboration] (2004)
  %``High resolution observations of the CMB power spectrum with ACBAR,''
\emph{  Astrophys.\ J.\ } {\bf 600} 32 %(2004)
(\emph{Preprint}   astro-ph/0212289)
  %%CITATION = ASTRO-PH/0212289;%%
\bibitem{EH97}
  Eisenstein D J and Hu W 1998
%  \emph{Baryonic Features in the Matter Transfer Function,}
\emph{
  Astrophys.\ J.\  }{\bf 496} 605
  (\emph{Preprint} astro-ph/9709112)
  %%CITATION = ASJOA,496,605;%%
\bibitem{inflation}
  Linde A 2008
%  \emph{Inflationary Cosmology,}
\emph{
  Lect.\ Notes Phys.\  }{\bf 738} 1
  (\emph{Preprint} 0705.0164 [hep-th]).
  %%CITATION = LNPHA,738,1;%%
\end{thebibliography}
\end{document}